\documentclass[journal=jctcce, manuscript=letter, layout=twocolumn]{achemso}
\setkeys{acs}{doi=true}

\usepackage{graphicx}
\usepackage{color}
\usepackage{amssymb}
\usepackage{amsmath}
\usepackage[utf8]{inputenc} 

\usepackage{verbatim}
\usepackage{siunitx}
\DeclareSIUnit\angstrom{\text{Å}}
\usepackage{listings}
\usepackage{hyperref}
\hypersetup{
    colorlinks=true, 
    linktoc=all,     
    linkcolor=blue,  
    citecolor=blue,
}

\usepackage{doi}

\usepackage{lipsum}  
\usepackage{abstract}

\title{Ensemble Knowledge Distillation for Machine Learning Interatomic Potentials}

\let\oldmaketitle\maketitle  
\let\maketitle\relax

\author{Sakib Matin}
\email{sakibmatin@gmail.com}
\affiliation{Theoretical Division, Los Alamos National Laboratory, Los Alamos, NM 87545, USA}

\author{Emily Shinkle}
\affiliation{Computer, Computational, and Statistical Sciences Division, Los Alamos National Laboratory, Los Alamos, NM 87545, USA}

\author{Yulia Pimonova}
\affiliation{Computer, Computational, and Statistical Sciences Division, Los Alamos National Laboratory, Los Alamos, NM 87545, USA}

\author{Galen T. Craven}
\affiliation{Theoretical Division, Los Alamos National Laboratory, Los Alamos, NM 87545, USA}

\author{Aleksandra Pachalieva}
\affiliation{Earth and Environmental Sciences Division, Los Alamos National Laboratory, Los Alamos, NM 87545, USA}

\author{Ying Wai Li}
\affiliation{Computer, Computational, and Statistical Sciences Division, Los Alamos National Laboratory, Los Alamos, NM 87545, USA}

\author{Kipton Barros}
\affiliation[lanlt]{Theoretical Division, Los Alamos National Laboratory, Los Alamos, NM 87545, USA}
\alsoaffiliation{Center for Nonlinear Studies, Los Alamos National Laboratory, Los Alamos, New Mexico 87545}

\author{Nicholas Lubbers}
\affiliation{Computer, Computational, and Statistical Sciences Division, Los Alamos National Laboratory, Los Alamos, NM 87545, USA}


\begin{document}

\twocolumn[
\oldmaketitle
\begin{onecolabstract}
The quality of machine learning interatomic potentials (MLIPs) strongly depends on the quantity of training data as well as the quantum chemistry (QC) level of theory used.
Datasets generated with high-fidelity QC methods are typically restricted to small molecules and may be missing energy gradients, which make it difficult to train accurate MLIPs.
We present an ensemble knowledge distillation (EKD) method to improve MLIP accuracy when trained to energy-only datasets. First, multiple teacher models are trained to QC energies and then generate atomic forces for all configurations in the dataset. Next, the student MLIP is trained to both QC energies and to ensemble-averaged forces generated by the teacher models.
We apply this workflow on the ANI-1ccx dataset 
where the configuration energies computed at the coupled cluster level of theory. 
The resulting student MLIPs achieve new state-of-the-art accuracy on the 
COMP6 benchmark and show improved stability for molecular dynamics simulations. 
\end{onecolabstract}
\vspace{1cm}
]

Machine learning models are a promising way to accelerate scientific simulations~\cite{unke2021machine, kulichenko2021rise, fedik2022extending, pyzer2022accelerating}. 
Machine learning interatomic potentials (MLIPs)~\cite{unke2021machine, kulichenko2021rise, fedik2022extending} can emulate the potential energy surface of atomistic systems at dramatically reduced costs compared to reference quantum chemistry (QC) algorithms. MLIPs~\cite{zuo2020performance, deringer2021gaussian,  kulichenko2021rise, unke2021machine, fedik2022extending, duval2023hitchhiker, allen2024learning} are typically trained on QC datasets to map from atomic coordinates and species to configuration energy. The atomic forces can be predicted by the trained MLIPs, using automatic differentiation. 
There has been tremendous progress in the field of MLIP development, especially in terms of designing more expressive architectures, including equivariant descriptors~\cite{batatia2022mace, batzner2022E3-equivariant, chigaev2023lightweight} and transformer-based architectures~\cite{liao2023equiformer}. Furthermore recent works have focused on generating larger training data sets~\cite{jain2013commentary, levine2025open}, which often utilize different active learning protocols~\cite{smith2018less, smith2021automated, van2022hyperactive,kulichenko2024data} ,as well as  scalable training~\cite{pasini2024scalable}. On the other hand, an under-explored area is the design of better training protocols, especially for low data regimes. Transfer learning~\cite{smith2019approaching} and multi-fidelity learning~\cite{kim2024data, messerly2025multifidelity} can be effective for small high fidelity datasets but they may require large amounts of data at a different level of theory to be successful. 

Generating the data used to train MLIPs is far more computationally expensive than performing the training itself~\cite{kulichenko2024data}. 
It is especially computationally difficult to generate training data using high fidelity QC methods, such as Coupled Cluster, Configuration-Interaction, 
Quantum Monte Carlo, etc, because computational cost typically explodes with the number of electrons~\cite{burke2012perspective}. Because of the much higher costs relative to density functional theory (DFT) calculations, datasets of high-fidelity QC are typically small, both in the size and number of molecular configurations. Furthermore, many high-fidelity QC codes provide only the total energy, but no gradients, which further hinders training of MLIPs~\cite{devereux2025force}.
Due to the great expense of obtaining the gold standard of accuracy in quantum chemistry data, methods to make the most of limited datasets are of great utility.

Recently, Knowledge Distillation (KD) for MLIPs~\cite{kelvinius2023accelerating, matin2025teacher} has been shown to be an effective training protocol for existing datasets, without expensive pre-training. 
In the prototypical KD workflow~\cite{hinton2015distilling}, a single teacher model generates auxiliary outputs that augment the training of a student model in order to enhance speed~\cite{yang2020distilling, kelvinius2023accelerating}, memory requirements~\cite{sanh2019distilbert}, and accuracy~\cite{furlanello2018born}. References~\citenum{chebotar2016distilling, furlanello2018born, asif2020ensemble} have shown that multiple teachers can train a single student model to improve performance in classification tasks. 
KD has been applied to MLIPs to accelerate molecular dynamics (MD) simulations, by using intermediate outputs (e.g., atomic energies)~\cite{matin2025teacher}, learned features~\cite{kelvinius2023accelerating}, and hessians of the energy~\cite{amin2025towards}. Similar teacher-student training has also been applied to materials structure for property prediction tasks~\cite{zhu2024addressing} and physics-constrained data augmentation~\cite{f2025improving}. 
In related works (see Refs.~\citenum{morrow2022indirect, gardner2023synthetic, gardner2024synthetic}), the teacher MLIP (trained on QC ground truth data) is used to generate synthetic data by running MD under different conditions. The student MLIPs is first trained to the synthetic data~\cite{morrow2022indirect, gardner2023synthetic}, and optionally fine-tuned on other QC ground truth data~\cite{gardner2024synthetic}. 
Existing studies have mostly focus on utilizing a single teacher and student workflow, even though using ensemble-averaged MLIPs 
in MD simulations has been shown to improve accuracy and stability~\cite{smith2021automated}. 
Only Ref.~\citenum{gong2025predictive} studied knowledge distillation with an ensemble five graph neural networks (GNNs) trained on energy, forces and stress, which were evaluated at DFT level of theory. A GNN was chosen at random to generate new trajectories, for which the ensemble was used to predict energy, forces and stress, and then the model was fine-tuned to the mean predictions~\cite{gong2025predictive}. 

\begin{figure*}[h]
    \centering
    \includegraphics[width=1.0\textwidth]{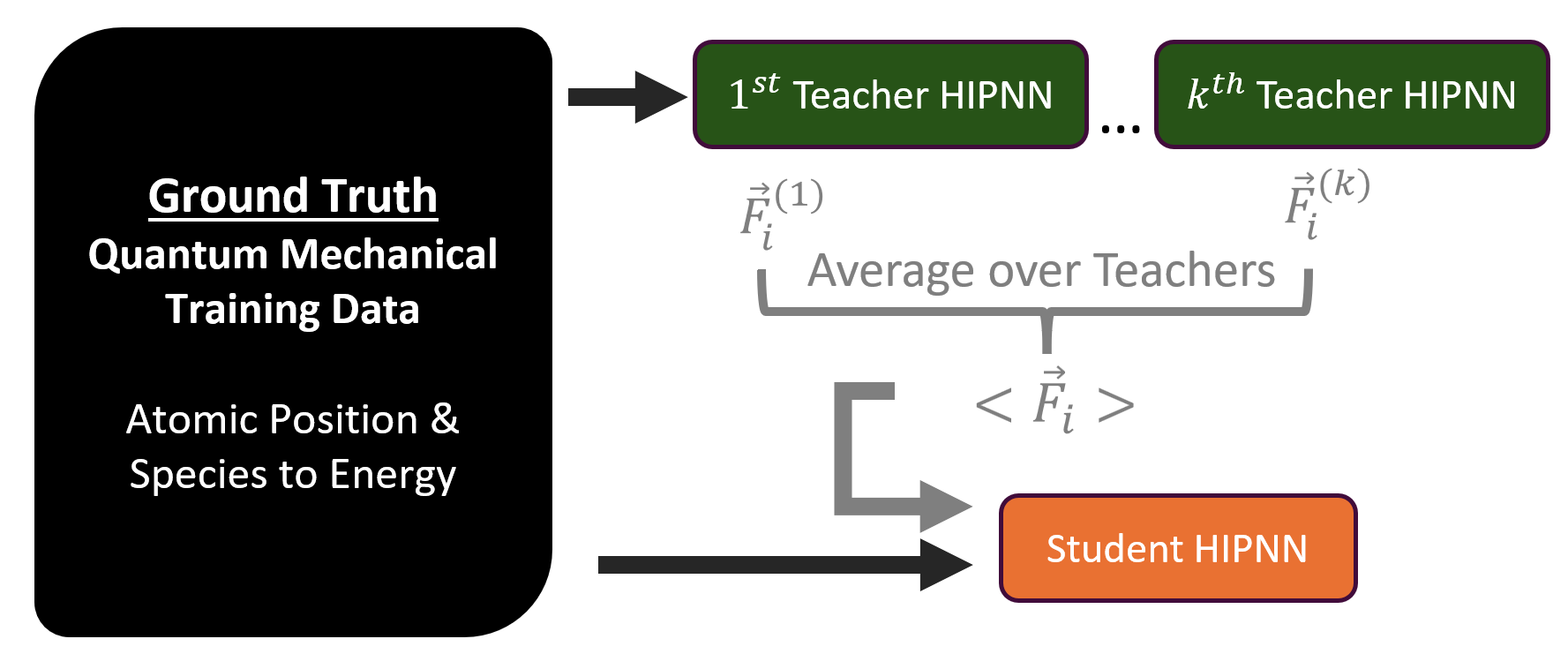}
    \caption{\textbf{Ensemble Knowledge Distillation  
    for Hierarchically Interacting Particle Neural Network (HIPNN)~\cite{lubbers2018hierarchical, chigaev2023lightweight}.} 
    The $k$ teacher models are trained on the 
    reference quantum chemistry calculation of the energy 
    and can generate forces (negative of the gradient of energy with respect to position). 
    These forces are averaged over the ensemble of teachers and augment the student HIPNN training to improve the accuracy and robustness. }
    \label{fig:workflow}
\end{figure*}

Distinct from previous publications~\cite{kelvinius2023accelerating, zhu2024addressing, gong2025predictive, matin2025teacher}, in our Ensemble Knowledge Distillation (EKD) for MLIPs workflow, a set of teacher MLIPs are trained on high fidelity data that only contains molecular energies. 
The trained teacher models can generate the atomic forces for all configurations in the dataset. 
The students are trained to the ground truth QC energies and the ensemble averaged forces from the teachers. Our workflow is outlined in Fig~\ref{fig:workflow}. We validate our EKD workflow on the ANI-1ccx dataset~\cite{smith2019approaching, smith2020ani} to show student models are more accurate and robust in our molecular dynamics tests than  direct training. Our workflow establishes a new state-of-the-art accuracy for the ANI-1ccx dataset.

In this letter, we apply the EKD method primarily to the
Hierarchically Interacting Particle Neural Network (HIPNN) architecture~\cite{lubbers2018hierarchical}, although this method can be readily applied to most existing MLIPs. HIPNN is a message-passing graph neural network~\cite{duval2023hitchhiker}, that can map atomic configurations to energy~\cite{lubbers2018hierarchical}, forces~\cite{chigaev2023lightweight}, and to various chemical properties~\cite{sifain2018discovering, magedov2021bond, li2024machine}. 
HIPNN uses one-hot encoding, based on the atomic number, to initially featurize the local atomic environments. 
Then the interaction layers allow for mixing of atomic environments between neighbors (within a local cutoff) via message passing to refine the initial features~\cite{lubbers2018hierarchical}. 
Using multiple interaction layers, $n_\mathrm{Int}>1$, implicitly accounts for some long-range physics~\cite{lubbers2018hierarchical, chigaev2023lightweight, duval2023hitchhiker}. 
Although HIPNN~\cite{lubbers2018hierarchical} originally utilized scalar pair-wise distances between neighbors,
the subsequent generalization---HIPNN with tensor sensitivity~\cite{chigaev2023lightweight, allen2025optimal}---utilizes higher order Cartesian tensor products of the displacement vectors between neighboring atoms to construct more informative many-body descriptors. 
The hyper-parameter $l_\mathrm{max}$ corresponds to the highest order tensor used in the model. For $\ell_\mathrm{max}>0$, the model predictions are sensitive to the angles between neighboring atoms. The $\ell_\mathrm{max}=0$ HIPNN model 
coincides with the model developed in original publication~\cite{lubbers2018hierarchical}.  
The atom layers (multi-layer perceptrons) predict the hierarchical contributions to the atomic energy, $\mathcal{\epsilon}_i$ 
which are then summed up to obtain the configuration, $E$. Automatic differentiation can be used to compute the forces on each atom $\boldsymbol{F}_i=-\boldsymbol{\nabla}_i E$. 
The hyper-parameters for the HIPNN models are given in 
Supporting Information.
~\ref{app:hyper}.

We validate our EKD workflow on the 

ANI-1ccx~\cite{smith2019approaching, smith2020ani} dataset, which consists of small organic molecules. The approximately $4.9 \times 10^5$ molecular configurations in this dataset span $\mathrm{C,H,N,O}$ elements.
The configurations have been down selected from the larger ANI-1x dataset (about 5 million configurations) using active learning~\cite{smith2019approaching}, 
which utilized ensemble disagreement as the uncertainty metric. 
The non-equilibrium geometries are generated via normal mode sampling and short MD trajectories~\cite{smith2018less}. The ANI-1ccx datasets include additional dihedral sampling of small molecules that is not present in the ANI-1x dataset. 
The dataset is available for download from Ref.~\citenum{smith2020ani}.
The configuration energies are computed at the coupled cluster with singles doubles, perturbative-triples, and complete basis set extrapolation [CCSD(T)/CBS] level of theory~\cite{riplinger2016sparse, smith2019approaching} using the ORCA software~\cite{neese2012orca}. The dataset has also been computed at the DFT level of theory using the $\omega \text{B} 97x$ functional, which is the same level of theory as the ANI-1x dataset~\cite{smith2018less}. 

The COMP6~\cite{smith2018less, smith2019approaching} is a challenging out-of-sample test for models trained to the ANI-1ccx and ANI-1x datasets. 
The configurations in the COMP6 are larger than the ANI-1x and ANI-1ccx training datasets, and provide a challenging extensibility test. 
For the `torsion'~\cite{sellers2017comparison} and `GDB 10-13' data subsets, the energy and conformer energy differences have been computed at the CCSD(T)/CBS level of theory and we refer to these as the CC-COMP6 dataset. The conformational energy $\Delta E$ is the energy difference between all conformers for a given molecule in the benchmark. The conformers with energies at least $100 \mathrm{kcal/mol}$ greater than the ground state are excluded, similar to the analysis in Ref.~\citenum{smith2019approaching}.

\begin{figure}
    \centering
    \includegraphics[]{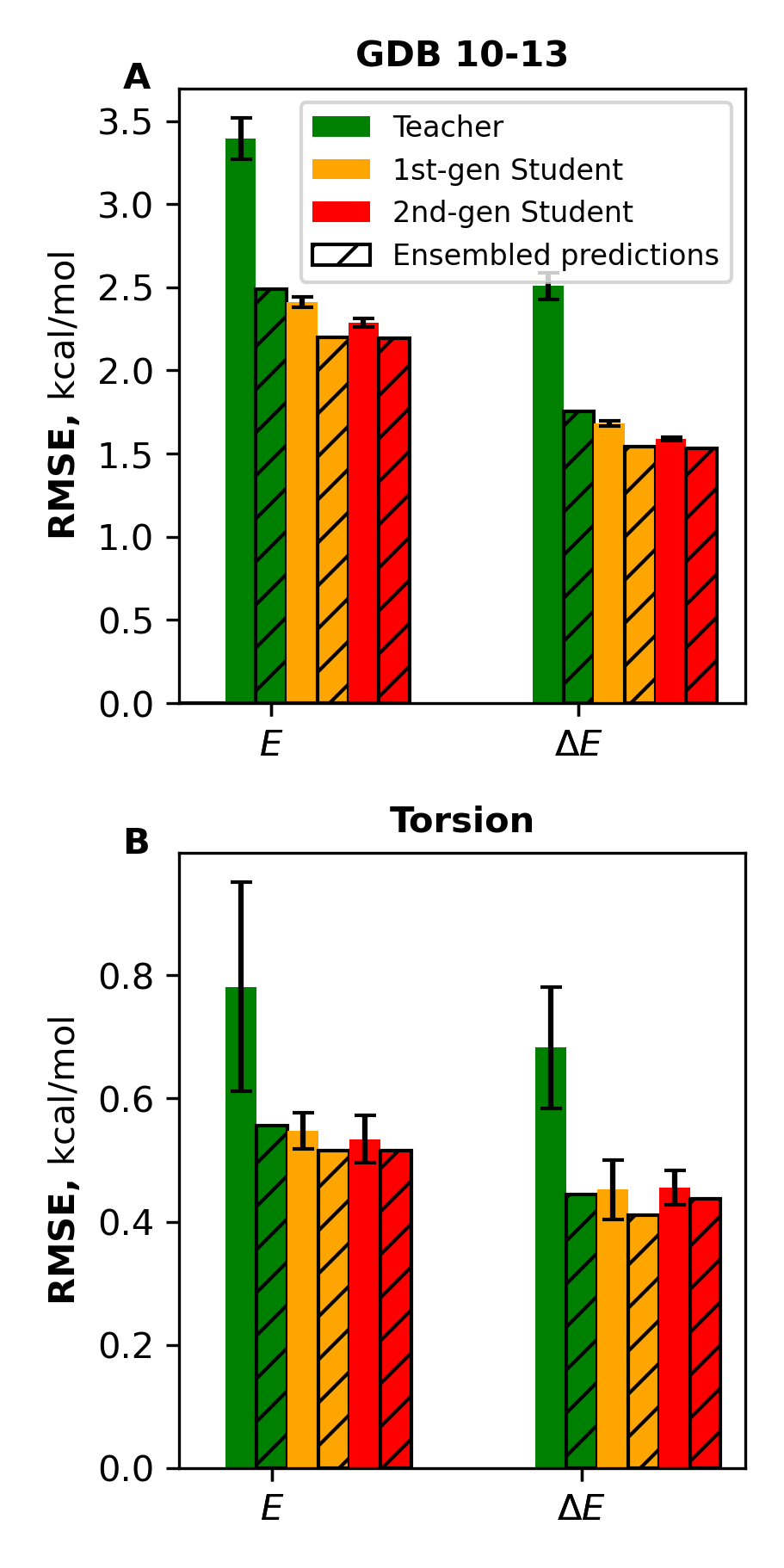}
    \caption{ \textbf{Student HIPNNs have lower root-mean-squared-errors (RMSE) for energy $E$ and conformer energy differences $\Delta E$ compared to the teacher models.} 
    The error bars correspond to the standard deviation measured across $8$ models which differ by random weight initializations and data splits.
    The out-of-sample test sets ``GDB 10-13" and ``Torsion''~\cite{sellers2017comparison} in panels A and B 
    are subsets of the CC-COMP6 dataset~\cite{smith2020ani}. 
    }
    \label{fig:comp6}
\end{figure}

We now outline the training procedure for our EKD method.
In the first step, we train an ensemble of eight  
teacher models on the QC dataset 
\begin{align}
    \mathfrak{D}:\left\{\boldsymbol{R}_i, Z_i\right\} \to \left\{E\right\},
\end{align}
which contains the atomic positions $\boldsymbol{R}_i$ and species $Z_i$ for each configuration and the corresponding energy $E$. 
The choice of $8$ teachers was motivated by ensemble knowledge distillation for image classification~\cite{allen2020towards}, which highlighted diminishing returns beyond $10$ neural networks. In the Supplementary Information~\ref{app:kd}, we show that using a single teacher model is less effective than using an ensemble of $8$. 
All $8$ teacher models have the same architecture, but are initialized with different random weights and different data splits. We train the HIPNN models using stochastic gradient descent. The loss function consists of both error and regularization terms. The error loss $\mathcal{L}_\mathrm{err}$ is the sum of the root-mean-squared error (RMSE) and mean-absolute error (MAE) losses. 
The regularization term consists of the $L_2$ norm of model weights $\mathcal{L}_{L_2}$, which is commonly added to loss functions to reduce over-fitting, and the hierarchicality term $\mathcal{L}_\mathrm{R}$~\cite{lubbers2018hierarchical}, which is specific to HIPNN. 

For the teacher models, the overall loss function is 
\begin{align}
    \bold{\mathcal{L}}_\mathrm{Teacher} &= w_E \mathcal{L}_\mathrm{err}(\hat{E},E) +  w_{L_2}\mathcal{L}_{L_2} + w_R \mathcal{L}_{R}. 
    \label{eq_loss_teacher}
\end{align}

Although the teacher models are trained only on molecular energies, they can predict the forces using automatic differentiation. The ensemble-averaged teacher forces for each atom,
\begin{align}
    \overline{\boldsymbol{F}}_i = \frac{1}{N} \sum^8_{T=1} \boldsymbol{F}^{(Teacher)}_i,
\end{align}
are used to construct the augmented dataset
\begin{align}
    \mathfrak{\tilde{D}}: \left\{\boldsymbol{R}_i, Z_i\right\} \to \left\{E,  \overline{\boldsymbol{F}}_i \right\}.
\end{align}
Note that the augmented dataset $\mathfrak{\tilde{D}}$ retains the same input configurations as $\mathfrak{D}$. 

The student models are trained to the augmented dataset with the loss function 
\begin{align}
\begin{split}
    \bold{\mathcal{L}}_\mathrm{Student} &= w_E \mathcal{L}_\mathrm{err}(\hat{E},E) + w_F \mathcal{L}_\mathrm{err}(\hat{\boldsymbol{F}_i},\overline{\boldsymbol{F}}_i) 
    \\ 
    &+ w_{L_2}\mathcal{L}_{L_2} + w_R \mathcal{L}_{R}. 
    \label{eq_loss_student}
\end{split}
\end{align}
We use a loss scheduler where the $w_\mathrm{F}$ is dynamically updated during training. The value of $w_\mathrm{F}$ is larger during the early stages and slowly decreases to a smaller value. The student model training benefits from the local information provided by the $\overline{\boldsymbol{F}}_i$ in the early stages of training, followed by refinement on the QC configuration energies at the final stages. Similar loss scheduler~\cite{vita2023data} 
strategies have shown improved accuracy when training MLIPs. The weights of the loss function and scheduler are listed in the Supplementary Information~\ref{app:loss_schedule}. The force predictions from an ensemble of the first generation student models can be used to train the second generation of student HIPNNs.

We apply the EKD workflow to HIPNNs trained on the ANI-1ccx data set.
The student HIPNNs achieve lower root-mean-squared error (RMSE) for energy $E$ and conformer energy differences $\Delta E$ in the out-of-sample CC-COMP6 benchmark compared to the set of teacher models in Fig.~\ref{fig:comp6}. The average error 
of the student HIPNNs is lower than the error of the ensembled predictions of teacher models.
The ensembled teachers are more accurate than single teachers, but they have slower MD speed and increased memory requirements. By using EKD, we get a single student that does just as well or even better than the ensembled teachers. Thus, we can get all the gains in accuracy from the ensembled teacher model without the computational disadvantages of ensembled models.

\begin{figure}
    \centering
    \includegraphics[]{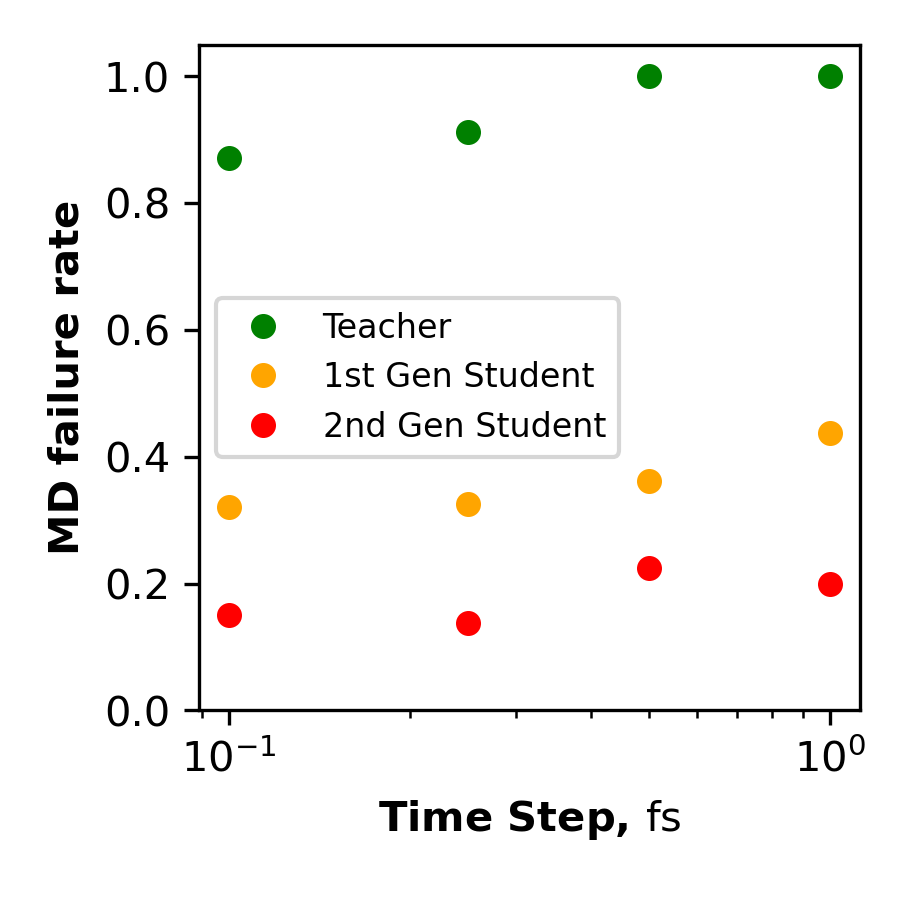}
    \caption{\textbf{MD simulations driven by student HIPNNs are more stable than teacher models.} The student and teachers are the same models as in Fig.~\ref{fig:comp6}. The MD simulations are performed with individual models, not ensembled predictions. Every data point is averaged over $8$ HIPNN models, where each MLIP is used to perform $10$ independent MD runs for a total of $80$ trajectories. }
    \label{fig:md-stability}
\end{figure}

We analyze the MD stability of all student and teacher HIPNNs. 
The Atomic Simulation Environment (ASE)~\cite{larsen2017atomic} is used to perform MD simulations
at constant number, energy and volume for different time step sizes
using 448 $\mathrm{CH_{3}ONO}$ molecules.  
The MD stability test in the condensed phase represents a difficult test of extensibility of the models because they were trained only on small gas phase clusters.
Each MD simulation is run for a maximum of $10 \mathrm{ps}$ unless it fails due to our close-contact criteria (smallest interatomic distance is smaller than $0.5\,\si{\angstrom}$). 
In Fig.~\ref{fig:md-stability}, we plot the fraction of MD runs that fail against the step size to show that the student HIPPNs, especially the second generation, are more robust than the teacher models across a range of time steps. We note that the second generation model's errors are comparable to the first generation, as seen in Fig.~\ref{fig:comp6}. Thus the EKD has improves the robustness of MLIPs beyond what is captured by the RMSE errors. 

\begin{figure}[h]
    \centering
    \includegraphics[]{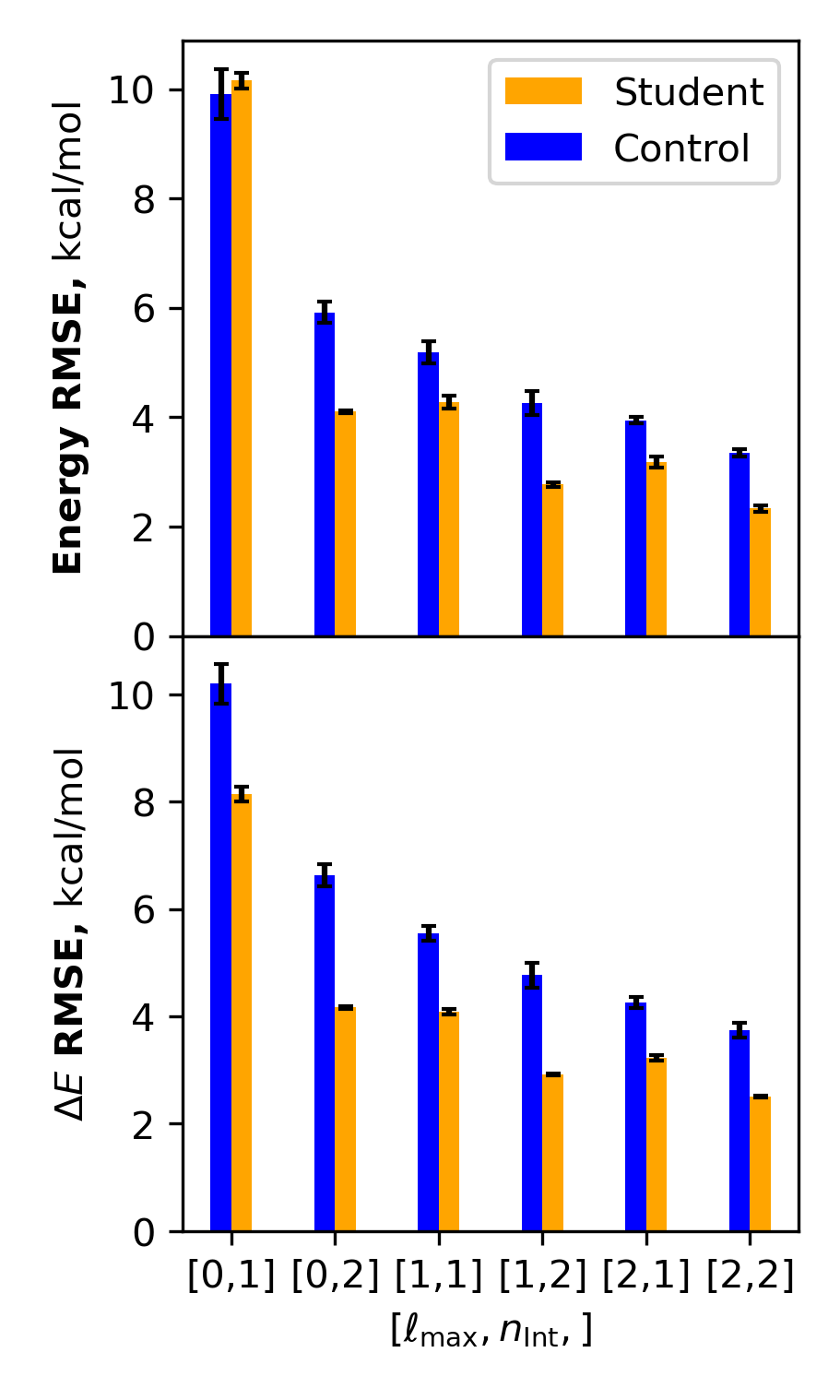}
    \caption{ \textbf{Student HIPNNs have lower root-mean-squared-errors (RMSE) on the out-of-sample CC-COMP6 benchmark than Control models.} HIPNN model capacity increases with increasing $n_\mathrm{int}$ and $\ell_\mathrm{max}$. Our ensemble knowledge distillation workflow is robust against the increasing capacity gap between teacher and student HIPNNs. 
    }
    \label{fig:cap-gap}
\end{figure}

To investigate how our EKD workflow is affected by the capacity gap between the teacher and student MLIPs, we vary the student models' 
number of interaction layers $n_\mathrm{Int}$ and tensor sensitivity order $\ell_\mathrm{max}$. Increasing $n_\mathrm{Int}$ and $l_\mathrm{max}$ means that descriptors are more sensitive to many-body angular information about neighboring atoms, i.e., the resulting models have greater capacity~\cite{chigaev2023lightweight}.
We fix the the width of the neural network, depth of the atom layers, and number of sensitivity functions, as well as the training hyper-parameters (number of epochs, learning rates, and optimizer) for all models considered. The control models  have the same $n_\mathrm{Int}$ and $\ell_\mathrm{max}$ as the students but they are trained only on the QC energies. We set $n_\mathrm{Int}=2, \ell_\mathrm{max}=2$ for the teacher HIPNNs. Figure~\ref{fig:cap-gap} shows that $E$ and $\Delta E$ RMSE of the student HIPNNs are consistently 10-30\% lower than the control models on the CC-COMP6 benchmark.

\begin{figure}
    \centering
    \includegraphics[]{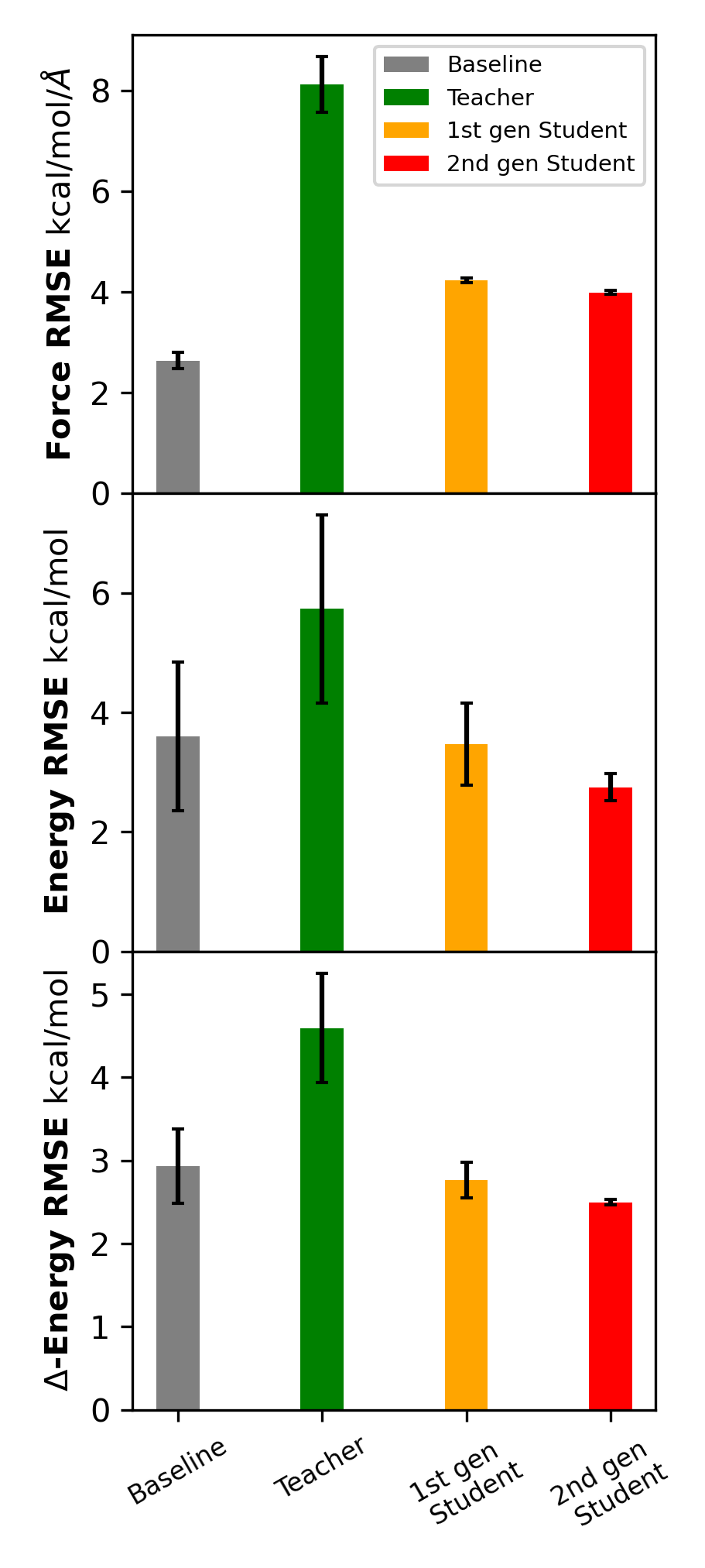}
    \caption{\textbf{Student models have lower force and energy errors than the teachers on the DFT-COMP6 benchmark}. 
    Only the ``Baseline'' models are trained to both the DFT energy and forces, and achieve the lowest force error.
    The student models achieve similar $E$ and $\Delta E$ errors as the baseline models.}
    \label{fig:dft-comp6}
\end{figure}

The accuracy of the forces of the student models in our EKD workflow is an important metric to analyze because of the strong correlation between the accuracy of forces and the accuracy of MD simulations.
Recall that the energy and forces for all configurations in ANI-1ccx dataset have been evaluated at the DFT level of theory using the $\omega \text{B} 97x$ functional~\cite{smith2020ani}, and we use this DFT-ANI-1ccx dataset only in Fig.~\ref{fig:dft-comp6}.
The teacher models are trained to the DFT ground truth energies. The first generation student models are trained to the ground truth DFT energies and the ensemble averaged forces from the teacher. The second generation student models are trained to the DFT energy and the ensemble averaged forces from the first generation students. Only the `baseline' model 
is trained to the DFT energy and forces. Figure~\ref{fig:dft-comp6} summarizes the energy and force errors for the DFT-COMP6 benchmark dataset. The student models have much lower force errors than the teachers but are higher than the `baseline' model which has access to the true energy and forces. The errors for $E$ and $\Delta E$ are comparable in the student and baseline models.

To summarize, we introduce the EKD framework for achieving higher MLIP accuracy when training to datasets that include energies but not forces. 
An ensemble of teacher MLIPs are trained on the QC energies and then used to generate forces for all the configurations in the dataset. Then, student HIPNNs are trained to the ground truth QC energies and the ensemble-averaged forces from the teachers.
The students exhibit up to $40 \%$ improvements in the out-of-sample CC-COMP6 benchmarks for the energy and conformer energy differences, as well as more stable MD simulations. To probe the accuracy of the students' forces, we apply the EKD to a dataset, where the energies and forces are computed at the DFT level of theory. The DFT forces are used only for testing, not for training, in the students, control and teachers. The student HIPNNs have lower errors with respect to the DFT forces when compared to the teachers. Additionally, the EKD workflow is effective even as the capacity gap between the teacher and student models grow. 
Although our workflow has the added cost of training an ensemble of teachers, it does not require any new expensive QC calculations beyond the original dataset needed to train the MLIPs, nor any exhaustive hyper-parameter tuning. Furthermore, our EKD workflow will be beneficial for reactive chemistry, where high fidelity QC methods are needed~\cite{johnson2008delocalization, hu2024training}. 
This is because the transition pathways generated with low fidelity methods, such as DFT, can show systematic deviations from high fidelity methods such as CCSDT~\cite{goerigk2011efficient, zhao2011density, vermeeren2022pericyclic, chamkin2024assessment} or CASPT~\cite{fedik2024challenges, hu2024training} due to excessive charge delocalization~\cite{johnson2008delocalization}. 

On a broader scope, our results are an important example of 
model-agnostic knowledge distillation for regression tasks using deep neural networks~\cite{amin2025towards}. Previous KD methods based on intermediate outputs have shown limited success~\cite{kelvinius2023accelerating, matin2025teacher} for regression, and feature matching KD~\cite{xu2020feature, kelvinius2023accelerating} approaches are dependent on the architectures of both the teachers and student models. This work paves the path towards 
model-agnostic KD methods, which will be relevant in constructing fast and accurate machine learning models.

\begin{acknowledgement}
We acknowledge support from the US DOE, Office of Science, Basic Energy Sciences, Chemical Sciences, Geosciences, and Biosciences Division under Triad National Security, LLC (“Triad”) contract Grant 89233218CNA000001 (FWP: LANLE3F2). Additionally, we acknowledge support from the Los Alamos National Laboratory (LANL) Directed Research and Development funds (LDRD). This research was performed in part at the Center for Nonlinear Studies (CNLS) at LANL. This research used resources provided by the Darwin testbed at LANL which is funded by the Computational Systems and Software Environments subprogram of LANL's Advanced Simulation and Computing program. 
\end{acknowledgement}

\begin{suppinfo}

\section{Knowledge Distillation using a single teacher \label{app:kd}}
We perform knowledge distillation with a single teacher model and a single student model for the Ani-1ccx dataset. The student model is trained to the ground truth energies and the forces from the teacher model. For the teacher model,  the ``GDB 10-13'' $E$ and $\Delta E$ RMSE are $3.31 \mathrm{kcal/mol}$, and $2.45 \mathrm{kcal/mol}$ respectively. The student HIPNN performs marginally better, and the  `GDB 10-13'' $E$ and $\Delta E$ RMSE are $2.93 \mathrm{kcal/mol}$, and $2.11 \mathrm{kcal/mol}$ respectively. We find that using only a single teacher model is less beneficial than using an ensemble, as seen in Fig.~\ref{fig:comp6}. 

\section{EKD across different MLIP architectures \label{app:ekd_ani-hipnn}}
\begin{figure}
    \centering
    \includegraphics[]{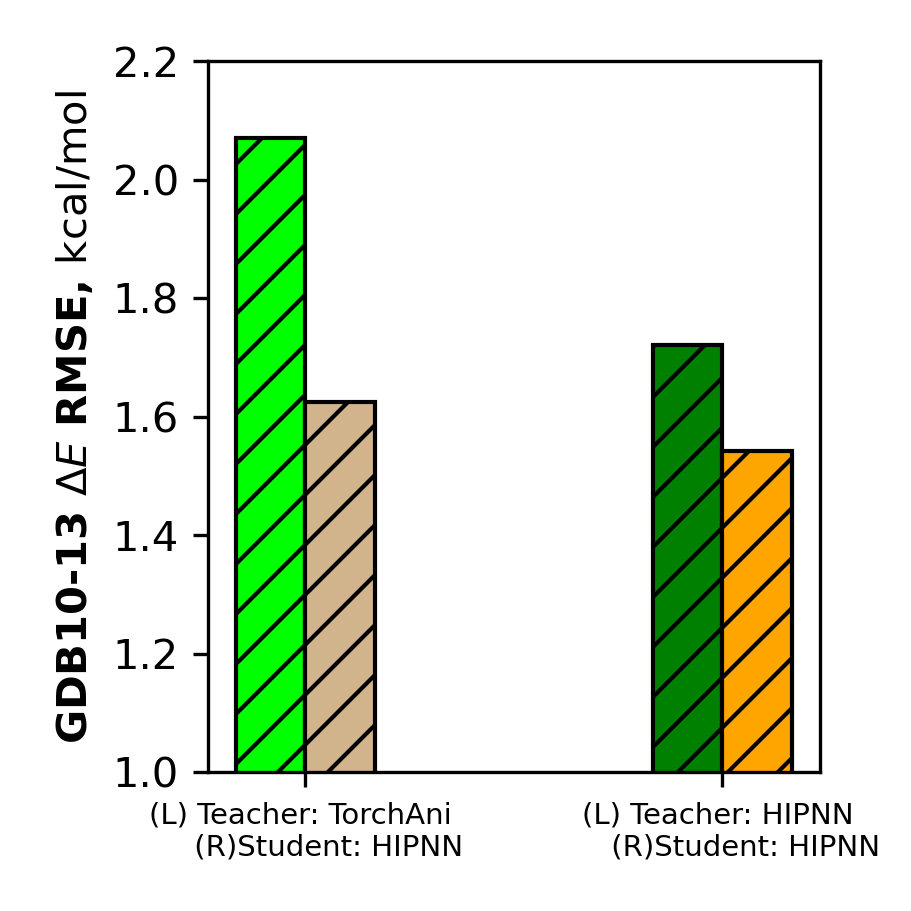}
    \caption{ \textbf{EKD is effective across different teacher student architectures.} The student HIPNN trained using forces from the TorchANI model has lower errors than the teacher HIPNN which was trained only to the QC energies. The diagonal hatch markings denote ensembled predictions.
    }
    \label{fig:ani-hip}
\end{figure}

We implement the EKD method with different MLIP architectures for teachers and students. We use TorchANI-1ccx MLIPs from Refs.~\citenum{aiqm2021torchani} and \citenum{gao2020torchani}, which were first pre-trained to the nearly 5 million configurations of the DFT ANI-1x dataset, and fine-tuned to the coupled cluster ANI-1ccx dataset~\cite{smith2019approaching}. We compare the $\Delta E$ RMSE of the ``GDB 10-13'' subset of the CC-COMP6 benchmark in Fig.~\ref{fig:ani-hip}. 
We find that the using a teacher torchANI model does benefit a student HIPNN model, compared to direct training (teacher HIPNNs). Ultimately, we find that using HIPNN MLIPs for both teacher and student gives lowest error, which we attribute to the fact the teacher HIPNNs have lower errors than the teacher TorchANI models.

\section{Training Details}
\subsection{HIPNN Hyper-Parameters\label{app:hyper}}
We use HIPNN models with $2$ interaction layers and  maximum tensor sensitivity order set at $l_\mathrm{max}=2$ for the teachers and the students HIPNNs, except in Sec.~\ref{fig:cap-gap}. All models use $4$ atom layers (feed-forward layers) with a width of $128$. The sensitivity functions, which parameterize the interaction layer, are characterized by radial cut-offs, namely, the soft maximum cutoff of $5.5~\si{\angstrom}$, and hard maximum cutoff of $6.5~\si{\angstrom}$ as well as a soft-min cutoff of $0.75~\si{\angstrom}$. All models use $20$ basis functions. The soft-min cut-off corresponds to the inner cut-off at very short distances. The hard maximum cut-off corresponds to the long distance cut-off. The soft maximum cutoff is set to a value smaller than the hard-dist cutoff to ensure a smooth truncation of the sensitivity functions. Note that the we are using the naming conventions for the hyper-parameters in the HIPNN GitHub Repository~\cite{lanl2021hippynn}, which differs slightly from the original HIPNN publication~\cite{lubbers2018hierarchical}.

\subsection{Loss Scheduler \label{app:loss_schedule}}
We summarize the weights corresponding to the loss function in Eq.~\ref{eq_loss_teacher}. 
The $W_\mathrm{E}=1, W_{L_2}=10^{-4},$ and $W_{R}=0.1$ is common to all models and remains static during training.  For the student models, we utilize a loss scheduler for the force term $w_F$ corresponding to Eq.~\ref{eq_loss_student}. During the early stages of training, the loss is heavily weighted to the auxiliary targets, namely, the ensemble averaged forces, and in the later stages the loss function is weighted more towards the QC energies. Our loss scheduler for $w_\mathrm{F}$ is summarized in Table~\ref{tabel:loss_scheduler}.
\begin{table}[h]
\small
  \caption{Loss Scheduler for force weight $w_\mathrm{F}$ for Student HIPNNs.}
  \label{tabel:loss_scheduler}
  \begin{tabular*}{0.48\textwidth}{@{\extracolsep{\fill}}ll}
    \hline
    Epoch & $w_F$ \\
    \hline
    1 & 10   \\
    20 & 9  \\
    40 & 8 \\ 
    60 & 7 \\ 
    80 & 6 \\ 
    100 & 5 \\ 
    120 & 4 \\ 
    140 & 3 \\ 
    160 & 2 \\ 
    180 & 1 \\ 
    200 & 0.5 \\
    \hline
  \end{tabular*}
\end{table}

\subsection{Optimizer\label{app:optim}}
We used the Adam Optimizer~\cite{kingma2017adam}, with an initial learning rate of $0.001$, which is halved with a patience of $15$ epochs. The termination patience is $30$ epochs. The maximum number of epochs is $400$. 

\end{suppinfo}

\bibliography{References}

\providecommand{\latin}[1]{#1}
\makeatletter
\providecommand{\doi}
  {\begingroup\let\do\@makeother\dospecials
  \catcode`\{=1 \catcode`\}=2 \doi@aux}
\providecommand{\doi@aux}[1]{\endgroup\texttt{#1}}
\makeatother
\providecommand*\mcitethebibliography{\thebibliography}
\csname @ifundefined\endcsname{endmcitethebibliography}  {\let\endmcitethebibliography\endthebibliography}{}
\begin{mcitethebibliography}{64}
\providecommand*\natexlab[1]{#1}
\providecommand*\mciteSetBstSublistMode[1]{}
\providecommand*\mciteSetBstMaxWidthForm[2]{}
\providecommand*\mciteBstWouldAddEndPuncttrue
  {\def\EndOfBibitem{\unskip.}}
\providecommand*\mciteBstWouldAddEndPunctfalse
  {\let\EndOfBibitem\relax}
\providecommand*\mciteSetBstMidEndSepPunct[3]{}
\providecommand*\mciteSetBstSublistLabelBeginEnd[3]{}
\providecommand*\EndOfBibitem{}
\mciteSetBstSublistMode{f}
\mciteSetBstMaxWidthForm{subitem}{(\alph{mcitesubitemcount})}
\mciteSetBstSublistLabelBeginEnd
  {\mcitemaxwidthsubitemform\space}
  {\relax}
  {\relax}

\bibitem[Unke \latin{et~al.}(2021)Unke, Chmiela, Sauceda, Gastegger, Poltavsky, Sch{\"u}tt, Tkatchenko, and M{\"u}ller]{unke2021machine}
Unke,~O.~T.; Chmiela,~S.; Sauceda,~H.~E.; Gastegger,~M.; Poltavsky,~I.; Sch{\"u}tt,~K.~T.; Tkatchenko,~A.; M{\"u}ller,~K.-R. {Machine Learning Force Fields}. \emph{{Chem. Rev.}} \textbf{2021}, \emph{121}, 10142--10186, DOI: \doi{10.1021/acs.chemrev.0c01111}\relax
\mciteBstWouldAddEndPuncttrue
\mciteSetBstMidEndSepPunct{\mcitedefaultmidpunct}
{\mcitedefaultendpunct}{\mcitedefaultseppunct}\relax
\EndOfBibitem
\bibitem[Kulichenko \latin{et~al.}(2021)Kulichenko, Smith, Nebgen, Li, Fedik, Boldyrev, Lubbers, Barros, and Tretiak]{kulichenko2021rise}
Kulichenko,~M.; Smith,~J.~S.; Nebgen,~B.; Li,~Y.~W.; Fedik,~N.; Boldyrev,~A.~I.; Lubbers,~N.; Barros,~K.; Tretiak,~S. The rise of neural networks for materials and chemical dynamics. \emph{{J. Phys. Chem. Lett.}} \textbf{2021}, \emph{12}, 6227--6243, DOI: \doi{10.1021/acs.jpclett.1c01357}\relax
\mciteBstWouldAddEndPuncttrue
\mciteSetBstMidEndSepPunct{\mcitedefaultmidpunct}
{\mcitedefaultendpunct}{\mcitedefaultseppunct}\relax
\EndOfBibitem
\bibitem[Fedik \latin{et~al.}(2022)Fedik, Zubatyuk, Kulichenko, Lubbers, Smith, Nebgen, Messerly, Li, Boldyrev, Barros, Isayev, and Tretiak]{fedik2022extending}
Fedik,~N.; Zubatyuk,~R.; Kulichenko,~M.; Lubbers,~N.; Smith,~J.~S.; Nebgen,~B.; Messerly,~R.; Li,~Y.~W.; Boldyrev,~A.~I.; Barros,~K.; Isayev,~O.; Tretiak,~S. {Extending machine learning beyond interatomic potentials for predicting molecular properties}. \emph{{Nat. Rev. Chem.}} \textbf{2022}, \emph{6}, 653--672, DOI: \doi{10.1038/s41570-022-00416-3}\relax
\mciteBstWouldAddEndPuncttrue
\mciteSetBstMidEndSepPunct{\mcitedefaultmidpunct}
{\mcitedefaultendpunct}{\mcitedefaultseppunct}\relax
\EndOfBibitem
\bibitem[Pyzer-Knapp \latin{et~al.}(2022)Pyzer-Knapp, Pitera, Staar, Takeda, Laino, Sanders, Sexton, Smith, and Curioni]{pyzer2022accelerating}
Pyzer-Knapp,~E.~O.; Pitera,~J.~W.; Staar,~P.~W.; Takeda,~S.; Laino,~T.; Sanders,~D.~P.; Sexton,~J.; Smith,~J.~R.; Curioni,~A. Accelerating materials discovery using artificial intelligence, high performance computing and robotics. \emph{npj Computational Materials} \textbf{2022}, \emph{8}, 84, DOI: \doi{10.1038/s41524-022-00765-z}\relax
\mciteBstWouldAddEndPuncttrue
\mciteSetBstMidEndSepPunct{\mcitedefaultmidpunct}
{\mcitedefaultendpunct}{\mcitedefaultseppunct}\relax
\EndOfBibitem
\bibitem[Zuo \latin{et~al.}(2020)Zuo, Chen, Li, Deng, Chen, Behler, Csányi, Shapeev, Thompson, Wood, and Ong]{zuo2020performance}
Zuo,~Y.; Chen,~C.; Li,~X.; Deng,~Z.; Chen,~Y.; Behler,~J.; Csányi,~G.; Shapeev,~A.~V.; Thompson,~A.~P.; Wood,~M.~A.; Ong,~S.~P. {Performance and Cost Assessment of Machine Learning Interatomic Potentials}. \emph{{The Journal of Physical Chemistry A}} \textbf{2020}, \emph{124}, 731--745, DOI: \doi{10.1021/acs.jpca.9b08723}, PMID: 31916773\relax
\mciteBstWouldAddEndPuncttrue
\mciteSetBstMidEndSepPunct{\mcitedefaultmidpunct}
{\mcitedefaultendpunct}{\mcitedefaultseppunct}\relax
\EndOfBibitem
\bibitem[Deringer \latin{et~al.}(2021)Deringer, Bart{\'o}k, Bernstein, Wilkins, Ceriotti, and Cs{\'a}nyi]{deringer2021gaussian}
Deringer,~V.~L.; Bart{\'o}k,~A.~P.; Bernstein,~N.; Wilkins,~D.~M.; Ceriotti,~M.; Cs{\'a}nyi,~G. {Gaussian process regression for materials and molecules}. \emph{{Chem. Rev.}} \textbf{2021}, \emph{121}, 10073--10141, DOI: \doi{10.1021/acs.chemrev.1c00022}\relax
\mciteBstWouldAddEndPuncttrue
\mciteSetBstMidEndSepPunct{\mcitedefaultmidpunct}
{\mcitedefaultendpunct}{\mcitedefaultseppunct}\relax
\EndOfBibitem
\bibitem[Duval \latin{et~al.}(2023)Duval, Mathis, Joshi, Schmidt, Miret, Malliaros, Cohen, Li{\`o}, Bengio, and Bronstein]{duval2023hitchhiker}
Duval,~A.; Mathis,~S.~V.; Joshi,~C.~K.; Schmidt,~V.; Miret,~S.; Malliaros,~F.~D.; Cohen,~T.; Li{\`o},~P.; Bengio,~Y.; Bronstein,~M. A Hitchhiker's Guide to Geometric GNNs for 3D Atomic Systems. \emph{arXiv preprint arXiv:2312.07511} \textbf{2023}, DOI: \doi{10.48550/arXiv.2312.07511}\relax
\mciteBstWouldAddEndPuncttrue
\mciteSetBstMidEndSepPunct{\mcitedefaultmidpunct}
{\mcitedefaultendpunct}{\mcitedefaultseppunct}\relax
\EndOfBibitem
\bibitem[Allen \latin{et~al.}(2024)Allen, Lubbers, Matin, Smith, Messerly, Tretiak, and Barros]{allen2024learning}
Allen,~A. E.~A.; Lubbers,~N.; Matin,~S.; Smith,~J.; Messerly,~R.; Tretiak,~S.; Barros,~K. Learning together: Towards foundation models for machine learning interatomic potentials with meta-learning. \emph{npj Computational Materials} \textbf{2024}, \emph{10}, 154, DOI: \doi{10.1038/s41524-024-01339-x}\relax
\mciteBstWouldAddEndPuncttrue
\mciteSetBstMidEndSepPunct{\mcitedefaultmidpunct}
{\mcitedefaultendpunct}{\mcitedefaultseppunct}\relax
\EndOfBibitem
\bibitem[Batatia \latin{et~al.}(2022)Batatia, Kovacs, Simm, Ortner, and Cs{\'a}nyi]{batatia2022mace}
Batatia,~I.; Kovacs,~D.~P.; Simm,~G.; Ortner,~C.; Cs{\'a}nyi,~G. {MACE: Higher order equivariant message passing neural networks for fast and accurate force fields}. \emph{{Advances in Neural Information Processing Systems}} \textbf{2022}, \emph{35}, 11423--11436, DOI: \doi{10.48550/arXiv.2206.07697}\relax
\mciteBstWouldAddEndPuncttrue
\mciteSetBstMidEndSepPunct{\mcitedefaultmidpunct}
{\mcitedefaultendpunct}{\mcitedefaultseppunct}\relax
\EndOfBibitem
\bibitem[Batzner \latin{et~al.}(2022)Batzner, Musaelian, Sun, Geiger, Mailoa, Kornbluth, Molinari, Smidt, and Kozinsky]{batzner2022E3-equivariant}
Batzner,~S.; Musaelian,~A.; Sun,~L.; Geiger,~M.; Mailoa,~J.~P.; Kornbluth,~M.; Molinari,~N.; Smidt,~T.~E.; Kozinsky,~B. {E(3)-equivariant graph neural networks for data-efficient and accurate interatomic potentials}. \emph{{Nat. Commun.}} \textbf{2022}, \emph{13}, 1--11, DOI: \doi{10.1038/s41467-022-29939-5}\relax
\mciteBstWouldAddEndPuncttrue
\mciteSetBstMidEndSepPunct{\mcitedefaultmidpunct}
{\mcitedefaultendpunct}{\mcitedefaultseppunct}\relax
\EndOfBibitem
\bibitem[Chigaev \latin{et~al.}(2023)Chigaev, Smith, Anaya, Nebgen, Bettencourt, Barros, and Lubbers]{chigaev2023lightweight}
Chigaev,~M.; Smith,~J.~S.; Anaya,~S.; Nebgen,~B.; Bettencourt,~M.; Barros,~K.; Lubbers,~N. {Lightweight and effective tensor sensitivity for atomistic neural networks}. \emph{The Journal of Chemical Physics} \textbf{2023}, \emph{158}, 184108, DOI: \doi{10.1063/5.0142127}\relax
\mciteBstWouldAddEndPuncttrue
\mciteSetBstMidEndSepPunct{\mcitedefaultmidpunct}
{\mcitedefaultendpunct}{\mcitedefaultseppunct}\relax
\EndOfBibitem
\bibitem[Liao and Smidt(2023)Liao, and Smidt]{liao2023equiformer}
Liao,~Y.-L.; Smidt,~T. Equiformer: Equivariant Graph Attention Transformer for 3D Atomistic Graphs. 2023; \url{https://arxiv.org/abs/2206.11990}\relax
\mciteBstWouldAddEndPuncttrue
\mciteSetBstMidEndSepPunct{\mcitedefaultmidpunct}
{\mcitedefaultendpunct}{\mcitedefaultseppunct}\relax
\EndOfBibitem
\bibitem[Jain \latin{et~al.}(2013)Jain, Ong, Hautier, Chen, Richards, Dacek, Cholia, Gunter, Skinner, Ceder, and Persson]{jain2013commentary}
Jain,~A.; Ong,~S.~P.; Hautier,~G.; Chen,~W.; Richards,~W.~D.; Dacek,~S.; Cholia,~S.; Gunter,~D.; Skinner,~D.; Ceder,~G.; Persson,~K.~A. {Commentary: The Materials Project: A materials genome approach to accelerating materials innovation}. \emph{APL Materials} \textbf{2013}, \emph{1}, 011002, DOI: \doi{10.1063/1.4812323}\relax
\mciteBstWouldAddEndPuncttrue
\mciteSetBstMidEndSepPunct{\mcitedefaultmidpunct}
{\mcitedefaultendpunct}{\mcitedefaultseppunct}\relax
\EndOfBibitem
\bibitem[Levine \latin{et~al.}(2025)Levine, Shuaibi, Spotte-Smith, Taylor, Hasyim, Michel, Batatia, Csányi, Dzamba, Eastman, Frey, Fu, Gharakhanyan, Krishnapriyan, Rackers, Raja, Rizvi, Rosen, Ulissi, Vargas, Zitnick, Blau, and Wood]{levine2025open}
Levine,~D.~S.; Shuaibi,~M.; Spotte-Smith,~E. W.~C.; Taylor,~M.~G.; Hasyim,~M.~R.; Michel,~K.; Batatia,~I.; Csányi,~G.; Dzamba,~M.; Eastman,~P.; Frey,~N.~C.; Fu,~X.; Gharakhanyan,~V.; Krishnapriyan,~A.~S.; Rackers,~J.~A.; Raja,~S.; Rizvi,~A.; Rosen,~A.~S.; Ulissi,~Z.; Vargas,~S.; Zitnick,~C.~L.; Blau,~S.~M.; Wood,~B.~M. The Open Molecules 2025 (OMol25) Dataset, Evaluations, and Models. 2025; \url{https://arxiv.org/abs/2505.08762}\relax
\mciteBstWouldAddEndPuncttrue
\mciteSetBstMidEndSepPunct{\mcitedefaultmidpunct}
{\mcitedefaultendpunct}{\mcitedefaultseppunct}\relax
\EndOfBibitem
\bibitem[Smith \latin{et~al.}(2018)Smith, Nebgen, Lubbers, Isayev, and Roitberg]{smith2018less}
Smith,~J.~S.; Nebgen,~B.; Lubbers,~N.; Isayev,~O.; Roitberg,~A.~E. {Less is more: Sampling chemical space with active learning}. \emph{{J. Chem. Phys.}} \textbf{2018}, \emph{148}, 241733, DOI: \doi{10.1063/1.5023802}\relax
\mciteBstWouldAddEndPuncttrue
\mciteSetBstMidEndSepPunct{\mcitedefaultmidpunct}
{\mcitedefaultendpunct}{\mcitedefaultseppunct}\relax
\EndOfBibitem
\bibitem[Smith \latin{et~al.}(2021)Smith, Nebgen, Mathew, Chen, Lubbers, Burakovsky, Tretiak, Nam, Germann, Fensin, and Barros]{smith2021automated}
Smith,~J.~S.; Nebgen,~B.; Mathew,~N.; Chen,~J.; Lubbers,~N.; Burakovsky,~L.; Tretiak,~S.; Nam,~H.~A.; Germann,~T.; Fensin,~S.; Barros,~K. {Automated discovery of a robust interatomic potential for aluminum}. \emph{{Nat. Commun.}} \textbf{2021}, \emph{12}, 1--13, DOI: \doi{10.1038/s41467-021-21376-0}\relax
\mciteBstWouldAddEndPuncttrue
\mciteSetBstMidEndSepPunct{\mcitedefaultmidpunct}
{\mcitedefaultendpunct}{\mcitedefaultseppunct}\relax
\EndOfBibitem
\bibitem[van~der Oord \latin{et~al.}(2022)van~der Oord, Sachs, Kov{\'a}cs, Ortner, and Cs{\'a}nyi]{van2022hyperactive}
van~der Oord,~C.; Sachs,~M.; Kov{\'a}cs,~D.~P.; Ortner,~C.; Cs{\'a}nyi,~G. {Hyperactive Learning (HAL) for Data-Driven Interatomic Potentials}. \emph{arXiv preprint arXiv:2210.04225} \textbf{2022}, DOI: \doi{10.48550/arXiv.2210.04225}\relax
\mciteBstWouldAddEndPuncttrue
\mciteSetBstMidEndSepPunct{\mcitedefaultmidpunct}
{\mcitedefaultendpunct}{\mcitedefaultseppunct}\relax
\EndOfBibitem
\bibitem[Kulichenko \latin{et~al.}(2024)Kulichenko, Nebgen, Lubbers, Smith, Barros, Allen, Habib, Shinkle, Fedik, Li, \latin{et~al.} others]{kulichenko2024data}
Kulichenko,~M.; Nebgen,~B.; Lubbers,~N.; Smith,~J.~S.; Barros,~K.; Allen,~A. E.~A.; Habib,~A.; Shinkle,~E.; Fedik,~N.; Li,~Y.~W.; others Data Generation for Machine Learning Interatomic Potentials and Beyond. \emph{Chemical Reviews} \textbf{2024}, DOI: \doi{10.1021/acs.chemrev.4c00572}\relax
\mciteBstWouldAddEndPuncttrue
\mciteSetBstMidEndSepPunct{\mcitedefaultmidpunct}
{\mcitedefaultendpunct}{\mcitedefaultseppunct}\relax
\EndOfBibitem
\bibitem[Pasini \latin{et~al.}(2024)Pasini, Choi, Mehta, Zhang, Rogers, Bae, Ibrahim, Aji, Schulz, Polo, and Balaprakash]{pasini2024scalable}
Pasini,~M.~L.; Choi,~J.~Y.; Mehta,~K.; Zhang,~P.; Rogers,~D.; Bae,~J.; Ibrahim,~K.~Z.; Aji,~A.~M.; Schulz,~K.~W.; Polo,~J.; Balaprakash,~P. Scalable Training of Trustworthy and Energy-Efficient Predictive Graph Foundation Models for Atomistic Materials Modeling: A Case Study with HydraGNN. 2024; \url{https://arxiv.org/abs/2406.12909}\relax
\mciteBstWouldAddEndPuncttrue
\mciteSetBstMidEndSepPunct{\mcitedefaultmidpunct}
{\mcitedefaultendpunct}{\mcitedefaultseppunct}\relax
\EndOfBibitem
\bibitem[Smith \latin{et~al.}(2019)Smith, Nebgen, Zubatyuk, Lubbers, Devereux, Barros, Tretiak, Isayev, and Roitberg]{smith2019approaching}
Smith,~J.~S.; Nebgen,~B.~T.; Zubatyuk,~R.; Lubbers,~N.; Devereux,~C.; Barros,~K.; Tretiak,~S.; Isayev,~O.; Roitberg,~A.~E. {Approaching coupled cluster accuracy with a general-purpose neural network potential through transfer learning}. \emph{{Nat. Commun.}} \textbf{2019}, \emph{10}, 1--8, DOI: \doi{10.1038/s41467-019-10827-4}\relax
\mciteBstWouldAddEndPuncttrue
\mciteSetBstMidEndSepPunct{\mcitedefaultmidpunct}
{\mcitedefaultendpunct}{\mcitedefaultseppunct}\relax
\EndOfBibitem
\bibitem[Kim \latin{et~al.}(2024)Kim, Kim, Kim, Lee, Park, Kang, and Han]{kim2024data}
Kim,~J.; Kim,~J.; Kim,~J.; Lee,~J.; Park,~Y.; Kang,~Y.; Han,~S. Data-Efficient Multifidelity Training for High-Fidelity Machine Learning Interatomic Potentials. \emph{Journal of the American Chemical Society} \textbf{2024}, DOI: \doi{10.1021/jacs.4c14455}\relax
\mciteBstWouldAddEndPuncttrue
\mciteSetBstMidEndSepPunct{\mcitedefaultmidpunct}
{\mcitedefaultendpunct}{\mcitedefaultseppunct}\relax
\EndOfBibitem
\bibitem[Messerly \latin{et~al.}(2025)Messerly, Matin, Allen, Nebgen, Barros, Smith, Lubbers, and Messerly]{messerly2025multifidelity}
Messerly,~M.; Matin,~S.; Allen,~A. E.~A.; Nebgen,~B.; Barros,~K.; Smith,~J.~S.; Lubbers,~N.; Messerly,~R. Multi-fidelity learning for interatomic potentials: Low-level forces and high-level energies are all you need. 2025; \url{https://arxiv.org/abs/2505.01590}\relax
\mciteBstWouldAddEndPuncttrue
\mciteSetBstMidEndSepPunct{\mcitedefaultmidpunct}
{\mcitedefaultendpunct}{\mcitedefaultseppunct}\relax
\EndOfBibitem
\bibitem[Burke(2012)]{burke2012perspective}
Burke,~K. {Perspective on density functional theory}. \emph{{J. Chem. Phys.}} \textbf{2012}, \emph{136}, 150901, DOI: \doi{10.1063/1.4704546}\relax
\mciteBstWouldAddEndPuncttrue
\mciteSetBstMidEndSepPunct{\mcitedefaultmidpunct}
{\mcitedefaultendpunct}{\mcitedefaultseppunct}\relax
\EndOfBibitem
\bibitem[Devereux \latin{et~al.}(2025)Devereux, Yang, Mart{\'\i}, Z{\'a}dor, Eldred, and Najm]{devereux2025force}
Devereux,~C.; Yang,~Y.; Mart{\'\i},~C.; Z{\'a}dor,~J.; Eldred,~M.~S.; Najm,~H.~N. Force training neural network potential energy surface models. \emph{{International Journal of Chemical Kinetics}} \textbf{2025}, \emph{57}, 59--76, DOI: \doi{10.1002/kin.21759}\relax
\mciteBstWouldAddEndPuncttrue
\mciteSetBstMidEndSepPunct{\mcitedefaultmidpunct}
{\mcitedefaultendpunct}{\mcitedefaultseppunct}\relax
\EndOfBibitem
\bibitem[Kelvinius \latin{et~al.}(2023)Kelvinius, Georgiev, Toshev, and Gasteiger]{kelvinius2023accelerating}
Kelvinius,~F.~E.; Georgiev,~D.; Toshev,~A.~P.; Gasteiger,~J. Accelerating Molecular Graph Neural Networks via Knowledge Distillation. \emph{arXiv preprint arXiv:2306.14818} \textbf{2023}, DOI: \doi{10.48550/arXiv.2306.14818}\relax
\mciteBstWouldAddEndPuncttrue
\mciteSetBstMidEndSepPunct{\mcitedefaultmidpunct}
{\mcitedefaultendpunct}{\mcitedefaultseppunct}\relax
\EndOfBibitem
\bibitem[Matin \latin{et~al.}(2025)Matin, Allen, Shinkle, Pachalieva, Craven, Nebgen, Smith, Messerly, Li, Tretiak, Barros, and Lubbers]{matin2025teacher}
Matin,~S.; Allen,~A.; Shinkle,~E.; Pachalieva,~A.; Craven,~G.~T.; Nebgen,~B.; Smith,~J.; Messerly,~R.; Li,~Y.~W.; Tretiak,~S.; Barros,~K.; Lubbers,~N. Teacher-student training improves accuracy and efficiency of machine learning inter-atomic potentials. 2025; \url{https://arxiv.org/abs/2502.05379}\relax
\mciteBstWouldAddEndPuncttrue
\mciteSetBstMidEndSepPunct{\mcitedefaultmidpunct}
{\mcitedefaultendpunct}{\mcitedefaultseppunct}\relax
\EndOfBibitem
\bibitem[Hinton \latin{et~al.}(2015)Hinton, Vinyals, and Dean]{hinton2015distilling}
Hinton,~G.; Vinyals,~O.; Dean,~J. Distilling the Knowledge in a Neural Network. \textbf{2015}, DOI: \doi{10.48550/arXiv.1503.02531}\relax
\mciteBstWouldAddEndPuncttrue
\mciteSetBstMidEndSepPunct{\mcitedefaultmidpunct}
{\mcitedefaultendpunct}{\mcitedefaultseppunct}\relax
\EndOfBibitem
\bibitem[Yang \latin{et~al.}(2020)Yang, Qiu, Song, Tao, and Wang]{yang2020distilling}
Yang,~Y.; Qiu,~J.; Song,~M.; Tao,~D.; Wang,~X. {Distilling Knowledge From Graph Convolutional Networks}. Proceedings of the IEEE/CVF Conference on Computer Vision and Pattern Recognition. 2020; pp 7074--7083, DOI: \doi{10.1109/CVPR42600.2020.00710}\relax
\mciteBstWouldAddEndPuncttrue
\mciteSetBstMidEndSepPunct{\mcitedefaultmidpunct}
{\mcitedefaultendpunct}{\mcitedefaultseppunct}\relax
\EndOfBibitem
\bibitem[Sanh \latin{et~al.}(2019)Sanh, Debut, Chaumond, and Wolf]{sanh2019distilbert}
Sanh,~V.; Debut,~L.; Chaumond,~J.; Wolf,~T. DistilBERT, a distilled version of BERT: smaller, faster, cheaper and lighter. \emph{arXiv preprint arXiv:1910.01108} \textbf{2019}, DOI: \doi{10.48550/arXiv.1910.01108}\relax
\mciteBstWouldAddEndPuncttrue
\mciteSetBstMidEndSepPunct{\mcitedefaultmidpunct}
{\mcitedefaultendpunct}{\mcitedefaultseppunct}\relax
\EndOfBibitem
\bibitem[Furlanello \latin{et~al.}(2018)Furlanello, Lipton, Tschannen, Itti, and Anandkumar]{furlanello2018born}
Furlanello,~T.; Lipton,~Z.; Tschannen,~M.; Itti,~L.; Anandkumar,~A. {Born Again Neural Networks}. {International Conference on Machine Learning}. 2018; pp 1607--1616, DOI: \doi{10.48550/arXiv.1805.04770}\relax
\mciteBstWouldAddEndPuncttrue
\mciteSetBstMidEndSepPunct{\mcitedefaultmidpunct}
{\mcitedefaultendpunct}{\mcitedefaultseppunct}\relax
\EndOfBibitem
\bibitem[Chebotar and Waters(2016)Chebotar, and Waters]{chebotar2016distilling}
Chebotar,~Y.; Waters,~A. {Distilling knowledge from ensembles of neural networks for speech recognition}. Interspeech. 2016; pp 3439--3443, DOI: \doi{10.21437/Interspeech.2016-1190}\relax
\mciteBstWouldAddEndPuncttrue
\mciteSetBstMidEndSepPunct{\mcitedefaultmidpunct}
{\mcitedefaultendpunct}{\mcitedefaultseppunct}\relax
\EndOfBibitem
\bibitem[Asif \latin{et~al.}(2020)Asif, Tang, and Harrer]{asif2020ensemble}
Asif,~U.; Tang,~J.; Harrer,~S. \emph{ECAI 2020}; IOS Press, 2020; pp 953--960, DOI: \doi{10.48550/arXiv.1909.08097}\relax
\mciteBstWouldAddEndPuncttrue
\mciteSetBstMidEndSepPunct{\mcitedefaultmidpunct}
{\mcitedefaultendpunct}{\mcitedefaultseppunct}\relax
\EndOfBibitem
\bibitem[Amin \latin{et~al.}(2025)Amin, Raja, and Krishnapriyan]{amin2025towards}
Amin,~I.; Raja,~S.; Krishnapriyan,~A. Towards Fast, Specialized Machine Learning Force Fields: Distilling Foundation Models via Energy Hessians. \emph{arXiv preprint arXiv:2501.09009} \textbf{2025}, DOI: \doi{10.48550/arXiv.2501.09009}\relax
\mciteBstWouldAddEndPuncttrue
\mciteSetBstMidEndSepPunct{\mcitedefaultmidpunct}
{\mcitedefaultendpunct}{\mcitedefaultseppunct}\relax
\EndOfBibitem
\bibitem[Zhu \latin{et~al.}(2024)Zhu, Xin, Zheng, Wang, and Yang]{zhu2024addressing}
Zhu,~D.; Xin,~Z.; Zheng,~S.; Wang,~Y.; Yang,~X. {Addressing the Accuracy-Cost Trade-off in Material Property Prediction Using a Teacher-Student Strategy}. \emph{{Journal of Chemical Theory and Computation}} \textbf{2024}, DOI: \doi{10.1021/acs.jctc.4c00625}\relax
\mciteBstWouldAddEndPuncttrue
\mciteSetBstMidEndSepPunct{\mcitedefaultmidpunct}
{\mcitedefaultendpunct}{\mcitedefaultseppunct}\relax
\EndOfBibitem
\bibitem[F.~dos Santos \latin{et~al.}(2025)F.~dos Santos, Nebgen, Allen, Hamilton, Matin, Smith, and Messerly]{f2025improving}
F.~dos Santos,~L.~G.; Nebgen,~B.~T.; Allen,~A. E.~A.; Hamilton,~B.~W.; Matin,~S.; Smith,~J.~S.; Messerly,~R.~A. Improving Bond Dissociations of Reactive Machine Learning Potentials through Physics-Constrained Data Augmentation. \emph{Journal of Chemical Information and Modeling} \textbf{2025}, DOI: \doi{10.1021/acs.jcim.4c01847}\relax
\mciteBstWouldAddEndPuncttrue
\mciteSetBstMidEndSepPunct{\mcitedefaultmidpunct}
{\mcitedefaultendpunct}{\mcitedefaultseppunct}\relax
\EndOfBibitem
\bibitem[Morrow and Deringer(2022)Morrow, and Deringer]{morrow2022indirect}
Morrow,~J.~D.; Deringer,~V.~L. {Indirect learning and physically guided validation of interatomic potential models}. \emph{{The Journal of Chemical Physics}} \textbf{2022}, \emph{157}, 104105, DOI: \doi{10.1063/5.0099929}\relax
\mciteBstWouldAddEndPuncttrue
\mciteSetBstMidEndSepPunct{\mcitedefaultmidpunct}
{\mcitedefaultendpunct}{\mcitedefaultseppunct}\relax
\EndOfBibitem
\bibitem[Gardner \latin{et~al.}(2023)Gardner, Faure~Beaulieu, and Deringer]{gardner2023synthetic}
Gardner,~J. L.~A.; Faure~Beaulieu,~Z.; Deringer,~V.~L. {Synthetic data enable experiments in atomistic machine learning}. \emph{{Digital Discovery}} \textbf{2023}, DOI: \doi{10.1039/D2DD00137C}\relax
\mciteBstWouldAddEndPuncttrue
\mciteSetBstMidEndSepPunct{\mcitedefaultmidpunct}
{\mcitedefaultendpunct}{\mcitedefaultseppunct}\relax
\EndOfBibitem
\bibitem[Gardner \latin{et~al.}(2024)Gardner, Baker, and Deringer]{gardner2024synthetic}
Gardner,~J.~L.; Baker,~K.~T.; Deringer,~V.~L. Synthetic pre-training for neural-network interatomic potentials. \emph{Machine Learning: Science and Technology} \textbf{2024}, \emph{5}, 015003, DOI: \doi{10.1088/2632-2153/ad1626}\relax
\mciteBstWouldAddEndPuncttrue
\mciteSetBstMidEndSepPunct{\mcitedefaultmidpunct}
{\mcitedefaultendpunct}{\mcitedefaultseppunct}\relax
\EndOfBibitem
\bibitem[Gong \latin{et~al.}(2025)Gong, Zhang, Mu, Pu, Wang, Han, Yu, Chen, Zheng, Wang, \latin{et~al.} others]{gong2025predictive}
Gong,~S.; Zhang,~Y.; Mu,~Z.; Pu,~Z.; Wang,~H.; Han,~X.; Yu,~Z.; Chen,~M.; Zheng,~T.; Wang,~Z.; others A predictive machine learning force-field framework for liquid electrolyte development. \emph{Nature Machine Intelligence} \textbf{2025}, 1--10, DOI: \doi{10.1038/s42256-025-01009-7}\relax
\mciteBstWouldAddEndPuncttrue
\mciteSetBstMidEndSepPunct{\mcitedefaultmidpunct}
{\mcitedefaultendpunct}{\mcitedefaultseppunct}\relax
\EndOfBibitem
\bibitem[Smith \latin{et~al.}(2020)Smith, Zubatyuk, Nebgen, Lubbers, Barros, Roitberg, Isayev, and Tretiak]{smith2020ani}
Smith,~J.~S.; Zubatyuk,~R.; Nebgen,~B.; Lubbers,~N.; Barros,~K.; Roitberg,~A.~E.; Isayev,~O.; Tretiak,~S. The ANI-1ccx and ANI-1x data sets, coupled-cluster and density functional theory properties for molecules. \emph{Scientific data} \textbf{2020}, \emph{7}, 134, DOI: \doi{10.1038/s41597-020-0473-z}\relax
\mciteBstWouldAddEndPuncttrue
\mciteSetBstMidEndSepPunct{\mcitedefaultmidpunct}
{\mcitedefaultendpunct}{\mcitedefaultseppunct}\relax
\EndOfBibitem
\bibitem[Lubbers \latin{et~al.}(2018)Lubbers, Smith, and Barros]{lubbers2018hierarchical}
Lubbers,~N.; Smith,~J.~S.; Barros,~K. {Hierarchical modeling of molecular energies using a deep neural network}. \emph{{J. Chem. Phys.}} \textbf{2018}, \emph{148}, 241715, DOI: \doi{10.1063/1.5011181}\relax
\mciteBstWouldAddEndPuncttrue
\mciteSetBstMidEndSepPunct{\mcitedefaultmidpunct}
{\mcitedefaultendpunct}{\mcitedefaultseppunct}\relax
\EndOfBibitem
\bibitem[Sifain \latin{et~al.}(2018)Sifain, Lubbers, Nebgen, Smith, Lokhov, Isayev, Roitberg, Barros, and Tretiak]{sifain2018discovering}
Sifain,~A.~E.; Lubbers,~N.; Nebgen,~B.~T.; Smith,~J.~S.; Lokhov,~A.~Y.; Isayev,~O.; Roitberg,~A.~E.; Barros,~K.; Tretiak,~S. {Discovering a Transferable Charge Assignment Model Using Machine Learning}. \emph{{The Journal of Physical Chemistry Letters}} \textbf{2018}, \emph{9}, 4495--4501, DOI: \doi{10.1021/acs.jpclett.8b01939}\relax
\mciteBstWouldAddEndPuncttrue
\mciteSetBstMidEndSepPunct{\mcitedefaultmidpunct}
{\mcitedefaultendpunct}{\mcitedefaultseppunct}\relax
\EndOfBibitem
\bibitem[Magedov \latin{et~al.}(2021)Magedov, Koh, Malone, Lubbers, and Nebgen]{magedov2021bond}
Magedov,~S.; Koh,~C.; Malone,~W.; Lubbers,~N.; Nebgen,~B. Bond order predictions using deep neural networks. \emph{{J. Appl. Phys.}} \textbf{2021}, \emph{129}, 064701, DOI: \doi{10.1063/5.0016011}\relax
\mciteBstWouldAddEndPuncttrue
\mciteSetBstMidEndSepPunct{\mcitedefaultmidpunct}
{\mcitedefaultendpunct}{\mcitedefaultseppunct}\relax
\EndOfBibitem
\bibitem[Li \latin{et~al.}(2024)Li, Lubbers, Tretiak, Barros, and Zhang]{li2024machine}
Li,~X.; Lubbers,~N.; Tretiak,~S.; Barros,~K.; Zhang,~Y. Machine Learning Framework for Modeling Exciton Polaritons in Molecular Materials. \emph{Journal of Chemical Theory and Computation} \textbf{2024}, \emph{20}, 891--901, DOI: \doi{10.1021/acs.jctc.3c01068}\relax
\mciteBstWouldAddEndPuncttrue
\mciteSetBstMidEndSepPunct{\mcitedefaultmidpunct}
{\mcitedefaultendpunct}{\mcitedefaultseppunct}\relax
\EndOfBibitem
\bibitem[Allen \latin{et~al.}(2025)Allen, Shinkle, Bujack, and Lubbers]{allen2025optimal}
Allen,~A. E.~A.; Shinkle,~E.; Bujack,~R.; Lubbers,~N. Optimal Invariant Bases for Atomistic Machine Learning. 2025; \url{https://arxiv.org/abs/2503.23515}\relax
\mciteBstWouldAddEndPuncttrue
\mciteSetBstMidEndSepPunct{\mcitedefaultmidpunct}
{\mcitedefaultendpunct}{\mcitedefaultseppunct}\relax
\EndOfBibitem
\bibitem[Riplinger \latin{et~al.}(2016)Riplinger, Pinski, Becker, Valeev, and Neese]{riplinger2016sparse}
Riplinger,~C.; Pinski,~P.; Becker,~U.; Valeev,~E.~F.; Neese,~F. Sparse maps—A systematic infrastructure for reduced-scaling electronic structure methods. II. Linear scaling domain based pair natural orbital coupled cluster theory. \emph{{The Journal of Chemical Physics}} \textbf{2016}, \emph{144}, DOI: \doi{10.1063/1.4939030}\relax
\mciteBstWouldAddEndPuncttrue
\mciteSetBstMidEndSepPunct{\mcitedefaultmidpunct}
{\mcitedefaultendpunct}{\mcitedefaultseppunct}\relax
\EndOfBibitem
\bibitem[Neese(2012)]{neese2012orca}
Neese,~F. The ORCA program system. \emph{Wiley Interdisciplinary Reviews: Computational Molecular Science} \textbf{2012}, \emph{2}, 73--78, DOI: \doi{10.1002/wcms.81}\relax
\mciteBstWouldAddEndPuncttrue
\mciteSetBstMidEndSepPunct{\mcitedefaultmidpunct}
{\mcitedefaultendpunct}{\mcitedefaultseppunct}\relax
\EndOfBibitem
\bibitem[Sellers \latin{et~al.}(2017)Sellers, James, and Gobbi]{sellers2017comparison}
Sellers,~B.~D.; James,~N.~C.; Gobbi,~A. A comparison of quantum and molecular mechanical methods to estimate strain energy in druglike fragments. \emph{Journal of chemical information and modeling} \textbf{2017}, \emph{57}, 1265--1275, DOI: \doi{10.1021/acs.jcim.6b00614}\relax
\mciteBstWouldAddEndPuncttrue
\mciteSetBstMidEndSepPunct{\mcitedefaultmidpunct}
{\mcitedefaultendpunct}{\mcitedefaultseppunct}\relax
\EndOfBibitem
\bibitem[Allen-Zhu and Li(2023)Allen-Zhu, and Li]{allen2020towards}
Allen-Zhu,~Z.; Li,~Y. Towards Understanding Ensemble, Knowledge Distillation and Self-Distillation in Deep Learning. 2023; \url{https://arxiv.org/abs/2012.09816}\relax
\mciteBstWouldAddEndPuncttrue
\mciteSetBstMidEndSepPunct{\mcitedefaultmidpunct}
{\mcitedefaultendpunct}{\mcitedefaultseppunct}\relax
\EndOfBibitem
\bibitem[Vita and Schwalbe-Koda(2023)Vita, and Schwalbe-Koda]{vita2023data}
Vita,~J.~A.; Schwalbe-Koda,~D. Data efficiency and extrapolation trends in neural network interatomic potentials. \emph{Machine Learning: Science and Technology} \textbf{2023}, \emph{4}, 035031, DOI: \doi{10.1088/2632-2153/acf115}\relax
\mciteBstWouldAddEndPuncttrue
\mciteSetBstMidEndSepPunct{\mcitedefaultmidpunct}
{\mcitedefaultendpunct}{\mcitedefaultseppunct}\relax
\EndOfBibitem
\bibitem[Hjorth~Larsen \latin{et~al.}(2017)Hjorth~Larsen, Jørgen~Mortensen, Blomqvist, Castelli, Christensen, Dułak, Friis, Groves, Hammer, Hargus, Hermes, Jennings, Bjerre~Jensen, Kermode, Kitchin, Leonhard~Kolsbjerg, Kubal, Kaasbjerg, Lysgaard, Bergmann~Maronsson, Maxson, Olsen, Pastewka, Peterson, Rostgaard, Schiøtz, Schütt, Strange, Thygesen, Vegge, Vilhelmsen, Walter, Zeng, and Jacobsen]{larsen2017atomic}
Hjorth~Larsen,~A.; Jørgen~Mortensen,~J.; Blomqvist,~J.; Castelli,~I.~E.; Christensen,~R.; Dułak,~M.; Friis,~J.; Groves,~M.~N.; Hammer,~B.; Hargus,~C.; Hermes,~E.~D.; Jennings,~P.~C.; Bjerre~Jensen,~P.; Kermode,~J.; Kitchin,~J.~R.; Leonhard~Kolsbjerg,~E.; Kubal,~J.; Kaasbjerg,~K.; Lysgaard,~S.; Bergmann~Maronsson,~J.; Maxson,~T.; Olsen,~T.; Pastewka,~L.; Peterson,~A.; Rostgaard,~C.; Schiøtz,~J.; Schütt,~O.; Strange,~M.; Thygesen,~K.~S.; Vegge,~T.; Vilhelmsen,~L.; Walter,~M.; Zeng,~Z.; Jacobsen,~K.~W. {The atomic simulation environment—a Python library for working with atoms}. \emph{{J. Phys.: Condens.Matter}} \textbf{2017}, \emph{29}, 273002, DOI: \doi{10.1088/1361-648X/aa680e}\relax
\mciteBstWouldAddEndPuncttrue
\mciteSetBstMidEndSepPunct{\mcitedefaultmidpunct}
{\mcitedefaultendpunct}{\mcitedefaultseppunct}\relax
\EndOfBibitem
\bibitem[Johnson \latin{et~al.}(2008)Johnson, Mori-S{\'a}nchez, Cohen, and Yang]{johnson2008delocalization}
Johnson,~E.~R.; Mori-S{\'a}nchez,~P.; Cohen,~A.~J.; Yang,~W. Delocalization errors in density functionals and implications for main-group thermochemistry. \emph{The Journal of chemical physics} \textbf{2008}, \emph{129}, DOI: \doi{10.1063/1.3021474}\relax
\mciteBstWouldAddEndPuncttrue
\mciteSetBstMidEndSepPunct{\mcitedefaultmidpunct}
{\mcitedefaultendpunct}{\mcitedefaultseppunct}\relax
\EndOfBibitem
\bibitem[Hu \latin{et~al.}(2024)Hu, Gordon, Johanessen, Tan, and Goodpaster]{hu2024training}
Hu,~Q.; Gordon,~A.; Johanessen,~A.; Tan,~L.; Goodpaster,~J. Training Transferable Interatomic Neural Network Potentials for Reactive Chemistry: Improved Chemical Space Sampling. \textbf{2024}, \relax
\mciteBstWouldAddEndPunctfalse
\mciteSetBstMidEndSepPunct{\mcitedefaultmidpunct}
{}{\mcitedefaultseppunct}\relax
\EndOfBibitem
\bibitem[Goerigk and Grimme(2011)Goerigk, and Grimme]{goerigk2011efficient}
Goerigk,~L.; Grimme,~S. Efficient and Accurate Double-Hybrid-Meta-GGA Density Functionals: Evaluation with the Extended GMTKN30 Database for General Main Group Thermochemistry, Kinetics, and Noncovalent Interactions. \emph{Journal of chemical theory and computation} \textbf{2011}, \emph{7}, 291--309, DOI: \doi{10.1021/ct100466k}\relax
\mciteBstWouldAddEndPuncttrue
\mciteSetBstMidEndSepPunct{\mcitedefaultmidpunct}
{\mcitedefaultendpunct}{\mcitedefaultseppunct}\relax
\EndOfBibitem
\bibitem[Zhao and Truhlar(2011)Zhao, and Truhlar]{zhao2011density}
Zhao,~Y.; Truhlar,~D.~G. Density functional theory for reaction energies: test of meta and hybrid meta functionals, range-separated functionals, and other high-performance functionals. \emph{Journal of Chemical Theory and Computation} \textbf{2011}, \emph{7}, 669--676, DOI: \doi{10.1021/ct1006604}\relax
\mciteBstWouldAddEndPuncttrue
\mciteSetBstMidEndSepPunct{\mcitedefaultmidpunct}
{\mcitedefaultendpunct}{\mcitedefaultseppunct}\relax
\EndOfBibitem
\bibitem[Vermeeren \latin{et~al.}(2022)Vermeeren, Dalla~Tiezza, Wolf, Lahm, Allen, Schaefer, Hamlin, and Bickelhaupt]{vermeeren2022pericyclic}
Vermeeren,~P.; Dalla~Tiezza,~M.; Wolf,~M.~E.; Lahm,~M.~E.; Allen,~W.~D.; Schaefer,~H.~F.; Hamlin,~T.~A.; Bickelhaupt,~F.~M. Pericyclic reaction benchmarks: hierarchical computations targeting CCSDT (Q)/CBS and analysis of DFT performance. \emph{Physical Chemistry Chemical Physics} \textbf{2022}, \emph{24}, 18028--18042, DOI: \doi{10.1039/D2CP02234F}\relax
\mciteBstWouldAddEndPuncttrue
\mciteSetBstMidEndSepPunct{\mcitedefaultmidpunct}
{\mcitedefaultendpunct}{\mcitedefaultseppunct}\relax
\EndOfBibitem
\bibitem[Chamkin and Chamkina(2024)Chamkin, and Chamkina]{chamkin2024assessment}
Chamkin,~A.~A.; Chamkina,~E.~S. Assessment of the applicability of DFT methods to [Cp* Rh]-catalyzed hydrogen evolution processes. \emph{Journal of Computational Chemistry} \textbf{2024}, \emph{45}, 2624--2639, DOI: \doi{10.1002/jcc.27468}\relax
\mciteBstWouldAddEndPuncttrue
\mciteSetBstMidEndSepPunct{\mcitedefaultmidpunct}
{\mcitedefaultendpunct}{\mcitedefaultseppunct}\relax
\EndOfBibitem
\bibitem[Fedik \latin{et~al.}(2024)Fedik, Li, Lubbers, Nebgen, Tretiak, and Li]{fedik2024challenges}
Fedik,~N.; Li,~W.; Lubbers,~N.; Nebgen,~B.; Tretiak,~S.; Li,~Y.~W. Challenges and Opportunities for Machine Learning Potentials in Transition Path Sampling: Alanine Dipeptide and Azobenzene Studies. \textbf{2024}, DOI: \doi{10.26434/chemrxiv-2024-8w526-v2}\relax
\mciteBstWouldAddEndPuncttrue
\mciteSetBstMidEndSepPunct{\mcitedefaultmidpunct}
{\mcitedefaultendpunct}{\mcitedefaultseppunct}\relax
\EndOfBibitem
\bibitem[Xu \latin{et~al.}(2020)Xu, Rui, Li, and Gu]{xu2020feature}
Xu,~K.; Rui,~L.; Li,~Y.; Gu,~L. Feature normalized knowledge distillation for image classification. European conference on computer vision. 2020; pp 664--680, DOI: \doi{10.1007/978-3-030-58595-2_40}\relax
\mciteBstWouldAddEndPuncttrue
\mciteSetBstMidEndSepPunct{\mcitedefaultmidpunct}
{\mcitedefaultendpunct}{\mcitedefaultseppunct}\relax
\EndOfBibitem
\bibitem[aiq()]{aiqm2021torchani}
TorchANI: Accurate Neural Network Potential on PyTorch \url{https://github.com/aiqm/torchani}.\relax
\mciteBstWouldAddEndPunctfalse
\mciteSetBstMidEndSepPunct{\mcitedefaultmidpunct}
{}{\mcitedefaultseppunct}\relax
\EndOfBibitem
\bibitem[Gao \latin{et~al.}(2020)Gao, Ramezanghorbani, Isayev, Smith, and Roitberg]{gao2020torchani}
Gao,~X.; Ramezanghorbani,~F.; Isayev,~O.; Smith,~J.~S.; Roitberg,~A.~E. {TorchANI: a free and open source PyTorch-based deep learning implementation of the ANI neural network potentials}. \emph{Journal of Chemical Information and Modeling} \textbf{2020}, \emph{60}, 3408--3415, DOI: \doi{10.1021/acs.jcim.0c00451}\relax
\mciteBstWouldAddEndPuncttrue
\mciteSetBstMidEndSepPunct{\mcitedefaultmidpunct}
{\mcitedefaultendpunct}{\mcitedefaultseppunct}\relax
\EndOfBibitem
\bibitem[lan()]{lanl2021hippynn}
The hippynn package - a modular library for atomistic machine learning with PyTorch GitHub repository. \url{https://github.com/lanl/hippynn}.\relax
\mciteBstWouldAddEndPunctfalse
\mciteSetBstMidEndSepPunct{\mcitedefaultmidpunct}
{}{\mcitedefaultseppunct}\relax
\EndOfBibitem
\bibitem[Kingma and Ba(2017)Kingma, and Ba]{kingma2017adam}
Kingma,~D.~P.; Ba,~J. {Adam: A Method for Stochastic Optimization}. 2017; \url{https://arxiv.org/abs/1412.6980}\relax
\mciteBstWouldAddEndPuncttrue
\mciteSetBstMidEndSepPunct{\mcitedefaultmidpunct}
{\mcitedefaultendpunct}{\mcitedefaultseppunct}\relax
\EndOfBibitem
\end{mcitethebibliography}

\clearpage
\begin{figure}
    \centering
    \includegraphics[width=0.75\textwidth]{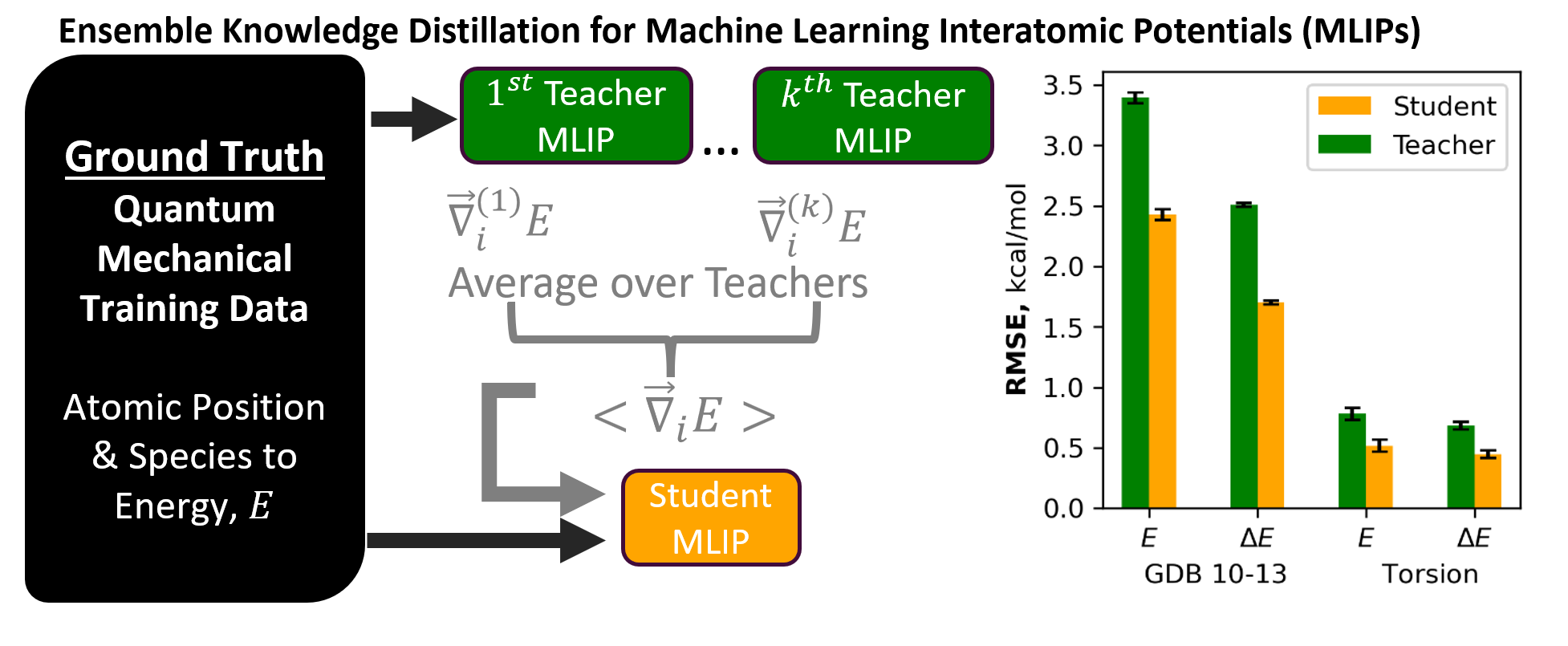}
    \label{fig:TOC}
\end{figure}

\end{document}